\documentclass{article}

\usepackage{PRIMEarxiv}

\usepackage[utf8]{inputenc} 
\usepackage[T1]{fontenc}    
\usepackage{hyperref}       
\usepackage{url}            
\usepackage{booktabs}       
\usepackage{amsfonts}       
\usepackage{nicefrac}       
\usepackage{microtype}      
\usepackage{lipsum}
\usepackage{fancyhdr}       
\usepackage{graphicx}       
\graphicspath{{media/}}     
\usepackage{mathtools}
\usepackage{caption}
\usepackage{graphics}
\usepackage{amsmath}
\usepackage{amssymb}
\usepackage{multirow}
\usepackage{cite}
\usepackage{array}
\usepackage{hhline}
\usepackage{float}
\usepackage{booktabs}

\usepackage{tabularx,booktabs}
\usepackage{makecell}
\newcolumntype{Y}{>{\raggedleft\arraybackslash}X}

\title{Reducing COVID-19 Cases and Deaths by Applying Blockchain in Vaccination Rollout Management
}

\author{
  Jorge Medina, Roberto Rojas-Cessa\\
  Affiliation \\
  New Jersey Institute of Technology \\
  Newark, NJ, USA\\
  \texttt{\{jorge.medina, rojas\}@njit.edu} \\
   \And
  Vatcharapan Umpaichitra\\
  Affiliation \\
  SUNY Downstate Health Sciences University \\
  Brooklyn, NY, USA\\
  \texttt{vatcharapan.umpaichitra@downstate.edu} \\
}

\begin{document}

\begin{titlepage}
	
	\scshape 
	
	\vspace*{\baselineskip} 
	
	\rule{\textwidth}{1.6pt}\vspace*{-\baselineskip}\vspace*{2pt} 
	\rule{\textwidth}{0.4pt} 
	
	\vspace{0.75\baselineskip} 
	
	{\textbf{Reducing COVID-19 Cases and Deaths by Applying Blockchain in Vaccination Rollout Management}} 
	
	\vspace{0.75\baselineskip} 
	
	\rule{\textwidth}{0.4pt}\vspace*{-\baselineskip}\vspace{3.2pt} 
	\rule{\textwidth}{1.6pt} 
	
	\vspace{2\baselineskip} 
	
\begin{abstract}
	
\textit{Goal:} 
\textit{Because a fast vaccination rollout against coronavirus disease 2019 (COVID-19) is critical to restore daily life and avoid virus mutations, it is tempting to have a relaxed vaccination-administration management system. However, a robust management system can support the enforcement of preventive measures, and in turn, reduce incidence and deaths. Here, we model a trustable and reliable management system based on blockchain for vaccine distribution by extending the Susceptible-Exposed-Infected-Recovery (SEIR) model. The model includes prevention measures such as mask-wearing, social distance, vaccination rate, and vaccination efficiency. It also considers negative social behavior, such as violations of social distance and attempts of using illegitimate vaccination proofs.
By evaluating the model, we show that the proposed system can reduce up to 2.5 million cases and half a million deaths in the most demanding scenarios.}\\

\textit{\textit{Impact Statement:} The use of blockchain technology on the system managing vaccination distribution enables a reliable exercise of infection prevention measures and a reduction of COVID-19 incidence and the number of deaths during and after vaccination rollout.}
\end{abstract}

\vspace{1.0\baselineskip}

This paper has been peer-reviewed and published as:
\textit{J. Medina, R. Cessa-Rojas and V. Umpaichitra, "Reducing COVID-19 Cases and Deaths by Applying Blockchain in Vaccination Rollout Management," in IEEE Open Journal of Engineering in Medicine and Biology, vol. 2, pp. 249-255, 2021, doi: 10.1109/OJEMB.2021.3093774. Errata: R. Cessa-Rojas is corrected here as R. Rojas-Cessa}

\vspace{1.0\baselineskip}	

{\textbf{Authors:}}
\vspace{0.5\baselineskip}
	
{\textbf{Jorge Medina}\\}
\textit{New Jersey Institute of Technology}\\
\textit{Newark, NJ, USA}\\
\textit{E-mail: jorge.medina@njit.edu}
	
\vspace{0.5\baselineskip}
{\textbf{Roberto Rojas-Cessa}\\}
\textit{New Jersey Institute of Technology}\\
\textit{Newark, NJ, USA}\\
\textit{E-mail: rojas@njit.edu}
	
\vspace{0.5\baselineskip}
{\textbf{Vatcharapan Umpaichitra}}
\textit{SUNY Downstate Health Sciences University}\\
\textit{Brooklyn, NY, USA}\\
\textit{E-mail: vatcharapan.umpaichitra@downstate.edu}
	
\vspace*{3\baselineskip} 

\vfill 

\vspace{0.3\baselineskip} 

\end{titlepage}

\maketitle

\begin{abstract}
\textit{Goal:} 
Because a fast vaccination rollout against coronavirus disease 2019 (COVID-19) is critical to restore daily life and avoid virus mutations, it is tempting to have a relaxed vaccination-administration management system. However, a robust management system can support the enforcement of preventive measures, and in turn, reduce incidence and deaths. Here, we model a trustable and reliable management system based on blockchain for vaccine distribution by extending the Susceptible-Exposed-Infected-Recovery (SEIR) model. The model includes prevention measures such as mask-wearing, social distance, vaccination rate, and vaccination efficiency. It also considers negative social behavior, such as violations of social distance and attempts of using illegitimate vaccination proofs.
By evaluating the model, we show that the proposed system can reduce up to 2.5 million cases and half a million deaths in the most demanding scenarios.\\\\

\textit{Impact Statement-} The use of blockchain technology on the system managing vaccination distribution enables a reliable exercise of infection prevention measures and a reduction of COVID-19 incidence and the number of deaths during and after vaccination rollout.
\end{abstract}


\section{Introduction}
With the ongoing coronavirus disease 2019 (COVID-19) pandemic, the world has been waiting for vaccines against the disease to counter its negative health and economic impacts that have affected everyone. COVID-19 is caused by the severe acute respiratory syndrome coronavirus 2 (SARS-CoV-2), which is highly contagious and has been fatal for millions of people~\cite{velavan2020covid,wu2020sars,pedersen2020sars, fathi2020time, fathi2021correlation}. COVID-19 is transmitted through droplets of saliva that are expelled when a person coughs, heavily breaths, talks, or even normally breathes. Detection of saliva is difficult as multiple of its components may need to be detected. Therefore, face-covering is considered a pivotal prevention measure as it provides a high degree of protection to both the wearer and the surrounding people. However, it is not 100\% protective. Therefore, there is hope that vaccinations will mitigate or even stop the spreading of this disease \cite{piraveenan2020optimal}.

Recently, the rollout of COVID-19 vaccines, while exercising social distancing and wearing face masks, is giving hope to the global population of restoring normal life. As more of the world population is now getting vaccinated, the management of the rollout and verification of vaccination records require not only accurate bookkeeping but also ubiquitous and secure access while maintaining user privacy. Accurate and accessible vaccination records are critical to provide an effective vaccination rollout and to support the execution of preventive measures, such as social distance~\cite{wu2020sars, krueger2021risk}.

COVID-19 vaccines from different manufacturers are being administered to the world population~\cite{vaccines2020randomized, burki2020russian,knoll2021oxford} aiming to vaccinate people rapidly and thus to leverage restoring normalcy, minimize health and economic damage, and reduce virus mutation opportunity~\cite{priesemann2021action}. The speedy distribution has to be carefully managed to avoid suboptimal benefits. However, the desirable synchronization and coordination of electronic vaccination records may be difficult to achieve in such a large vaccine distribution as it requires compatibility and administrative agreements among vaccination sites and upper administrators. Furthermore, the lack of secure processes to record vaccination events may make it difficult to enforce prevention measures at the rollout time and in the future. Vaccine verification could help to quickly restore daily social activities and traveling, as it may allow us to identify the susceptible and the potential source of infection. This verification requires reliable vaccination records that are secure, private, and accessible.

Blockchain has been proposed as a technology that can satisfy the security and reliability of vaccination records\cite{mcghin2019blockchain,ahmad2020blockchain,bansal2020optimizing,eisenstadt2020covid, hasan2020blockchain}. Beyond the technological advantages, a management system leveraged by blockchain may encourage people to get vaccinated and ultimately reduce the number of COVID-19 cases and deaths. In such approach, electronic immunization records are tied to verifiable blockchain transactions. Despite many recent proposals of blockchain-based vaccination certification approaches to store verifiable vaccination records, their impact on limiting or stopping the incidence of COVID-19 is still unknown.

To address this issue, we model the incidence and vaccination rollout by extending two Susceptible-Exposed-Infected-Recovered (SEIR) models. The SEIR-Vaccination (SEIR-V) model includes a vaccinated compartment using a conventional vaccination rollout campaign, and the SEIR-Vaccination-Blockchain (SEIR-VB) model uses blockchain to perform vaccination verification. With SEIR-VB, a policy of social distance can be encouraged by enabling vaccination verification as a passport to activities that may require reduced social distance. This paper shows that blockchain not only facilitates the management of vaccinations but also decreases the incidence of COVID-19 cases and deaths as compared to a system without blockchain. 

\section{MATERIALS AND METHODS}
Developed as a secure ledger in digital cryptocurrencies \cite{nakamoto2019bitcoin}, blockchain is an immutable distributed ledger technology that records verifiable transactional data~\cite{crosby2016blockchain}. Its immutability feature makes it applicable to a variety of healthcare systems~\cite{angraal2017blockchain, engelhardt2017hitching,linn2016blockchain}. A blockchain ledger consists of a cryptographically secured chain of chronologically ordered blocks of transactions. This ledger is distributed across multiple peers to ensure data availability and resiliency against failures and attacks~\cite{8643084}.

Bansal et al.~\cite{bansal2020optimizing} proposed to use blockchain as a repository of test results and vaccination records as verifiable immunization certificates. Hasan et at.~\cite{hasan2020blockchain} proposed a blockchain-based solution to securely store digital medical passports with immunity certificates for COVID-19. Eisenstadt et al.~\cite{eisenstadt2020covid} developed a prototype of a decentralized blockchain-based mobile application to store portable COVID-19 vaccination certificates. Such certificates may be used to verify the susceptibility of a person to contract COVID-19 and of being a source of infection.

Blockchain can also be used as a decentralized tracking information system. For example, Marbouh et at.~\cite{marbouh2020blockchain} proposed a blockchain that leverages smart contracts to transparently and reliably consolidate statistical data on COVID-19 incidence. Nguyen et at.~\cite{nguyen2020blockchain} presented a conceptual architecture that combines blockchain and artificial intelligence to monitor and track the COVID-19 outbreak in real-time.

But Blockchain can find its main application in detecting events that actually occurred, such as a vaccination, vaccine production, or social distance neglects. Tseng et al.~\cite{tseng2018governance} and Yong et al.~\cite{yong2020intelligent} proposed a blockchain system to verify the legitimacy of drugs and vaccines, respectively. Blockchain has been also proposed to monitor the efficacy of administered vaccines and dosages and to detect possible secondary effects~\cite{nguyen2020blockchain}. However, these works have not addressed the question that justifies its adoption during the COVID-19 vaccination rollout, as how helpful blockchain would be in such a campaign by preventing cases and deaths. We address this question in this paper through the following models.

\subsection{Extended SEIR Models}
The SEIR model is used to evaluate the spreading of an infectious disease over time~\cite{li1995global}. It categorizes individuals of a population into different compartments. The susceptible compartment is the group of individuals without immunity to the infection. The exposed compartment is the group of individuals who have been in close contact with infected individuals, and the recovered compartment is the group of the individuals who have recovered from the disease and developed immunity.

The prevention measures that reduce the number of COVID-19 cases considered in the extended SEIR models are mask wearing and social distancing. Vaccination against COVID-19 is a measure considered in the model to reduce cases and to convert a susceptible individual into an immune individual. This conversion depends on the vaccine efficacy, the vaccination rate, and the time for individuals to develop immunity. The willingness of individuals to get vaccinated also affects the immunity rate. Particularly, the models consider symptomatic and asymptomatic COVID-19 infections. The models also consider individuals' behaviors that affect the efficacy of these measures, such as neglecting social distance and falsely claiming wellness and vaccination.

The daily rate of change in the number of susceptible individuals, given by (\ref{eq:1}), is the sum of unvaccinated individuals that get exposed to both symptomatic and asymptomatic infected individuals, based on the contagious rate ($\beta$); the susceptible individuals that get vaccinated based on both the vaccination rate ($v_r$), and the likelihood of vaccine willingness ($l_v$); and the vaccinated individuals that remain susceptible based on the vaccine efficacy ($v_e$). The contagious rate $\beta$ is defined as the ratio between the basic reproduction number ($R_0$), which is defined by the average number of secondary cases that an infected individual infects, and the infection period~\cite{delamater2019complexity}.

Both SEIR-V and SEIR-VB models are described by (\ref{eq:1}) to (\ref{eq:7}), which determine the daily number of individuals in the respective compartments. Table 1 describes the variables used in the models. $S_k$ and $I_k$ represent the susceptible and infected individuals in the SEIR-V ($k$=$w$) and SEIR-VB ($k$=$b$) models, respectively (\ref{subscript}). The daily number of exposed individuals, given in (\ref{eq:2}), accounts for the newly exposed individuals; the exposed and unvaccinated individuals who may start to develop symptoms after the incubation period ($1/\delta$ days); and the fraction of the total exposed individuals that are vaccinated and recover as determined by the vaccine efficacy. The daily number of infected individuals includes symptomatic and asymptomatic cases, and both having a similar viral load~\cite{lee2020clinical}. A case-fatality rate ($\alpha$) is defined as the ratio of the number of deaths to the total number of cases.

The daily number of infected symptomatic individuals in (\ref{eq:3}) accounts for the exposed and unvaccinated individuals who get COVID-19 with a probability of symptomatic infection $p_{is}$; the cases that recover after the infection period of $1/\gamma$ days; the asymptomatic individuals that develop COVID-19 with probability $p_{ip}$ after $1/\lambda$ days; and the critical cases that unfortunately become deaths after the critical infection period of $1/\alpha$ days. The daily number of asymptomatic individuals in (\ref{eq:4}) accounts for the portion of exposed and unvaccinated individuals who become asymptomatic; the cases that develop infection; and the cases that remain asymptomatic and develop immunity with or without vaccination after $1/v_{i}$ days and $1/\mu$ days, respectively.

The daily number of recovered individuals, determined by (\ref{eq:5}), is the accumulated number of the recovered cases and the vaccinated individuals per day. A portion of the vaccinated individuals develop immunity. The portion is defined by the vaccine efficiency. The daily number of immune individuals in (\ref{eq:6}) is a function of both the recently vaccinated individuals and the vaccinated individuals that develop immunity. The daily number of deaths in (\ref{eq:7}) is the accumulated number of critical cases that result in deaths. 
\begin{align}
    & \frac{dS(t)}{dt} = -\frac{\beta S_{k}I_{k}(1-l_{v}v_{r})(I_{s}(t) + I_{a}(t))S(t)}{N} - l_{v}v_{r}S(t) + (1-v_{e})V(t)~\label{eq:1}\\\nonumber\\
    & \frac{dE(t)}{dt} = \frac{\beta S_{k}I_{k}(1-l_{v}v_{r})(I_{s}(t) + I_{a}(t))S(t)}{N}-\delta (1-l_{v}v_{r})E(t)-l_{v}v_{r}E(t)\label{eq:2}\\\nonumber\\
    & \frac{dI_{s}(t)}{dt} = \delta p_{is}(1-l_{v}v_{r})E(t) -\gamma(1-\alpha)I_{s}(t) +\lambda p_{ip}(1-l_{v}v_{r})I_{a}(t)-\rho \alpha I_{s}(t)~\label{eq:3}\\\nonumber\\
    & \frac{dI_{a}(t)}{dt} = \delta (1-p_{is})(1-l_{v}v_{r})E(t) -l_{v}v_{r}I_{a}(t) -\lambda p_{ip}(1-l_{v}v_{r})I_{a}(t) -\mu(1-p_{ip})(1-l_{v}v_{r})I_{a}(t) ~\label{eq:4}\\\nonumber\\
    &  \frac{dR(t)}{dt} = \gamma(1-\alpha)I_{s}(t)+\mu(1-p_{ip})(1-l_{v}v_{r})I_{a}(t) + v_{e}v_{i}V(t)\label{eq:5}\\\nonumber\\
    &  \frac{dV(t)}{dt} = (l_{v}v_{r})[S(t)+E(t)+I_{a}(t)] - (1-v_{e})V(t)  - v_{e}v_{i}V(t)\label{eq:6}\\\nonumber\\
    &\frac{dD(t)}{dt} = \rho \alpha I_{s}(t)~\label{eq:7}\\
    \text{Where:}\nonumber\\
    &k =
    \begin{cases}
      w & \textit{SEIR-V model.}\\
      b & \textit{SEIR-VB model.}
    \end{cases}~\label{subscript}
\end{align}

Then, the $S_k$ and $I_k$ for the SEIR-V and SEIR-VB models are:

\begin{align}
&S_{w} = (1-p_{m}m_{e})[\,p_{fv}+p_{sn}(1-p_{fv})\,]\label{eq:sw}\\
&I_{w} = (1-p_{m}m_{e})\{\,(1-p_{a})[\,p_{fv}+p_{sn}(1-p_{fv})\,] + p_{a}[\,p_{si}(1-p_{ci})+ p_{ci}\,]\}~\label{eq:iw}\\\nonumber\\
&S_{b} = (1-p_{m}m_{e})[\,p_{sn}(1-p_{fv})\,]~\label{eq:sb}\\
&I_{b} = (1-p_{m}m_{e})\{\,(1-p_{a})[\,p_{sn}(1-p_{fv})\,] + p_{a}[\,p_{si}(1-p_{ci})\,]\}~\label{eq:ib}
\end{align}

\begin{figure}[ht]
	\centering
	\includegraphics[width=\linewidth,trim=5 5 5 5,clip]{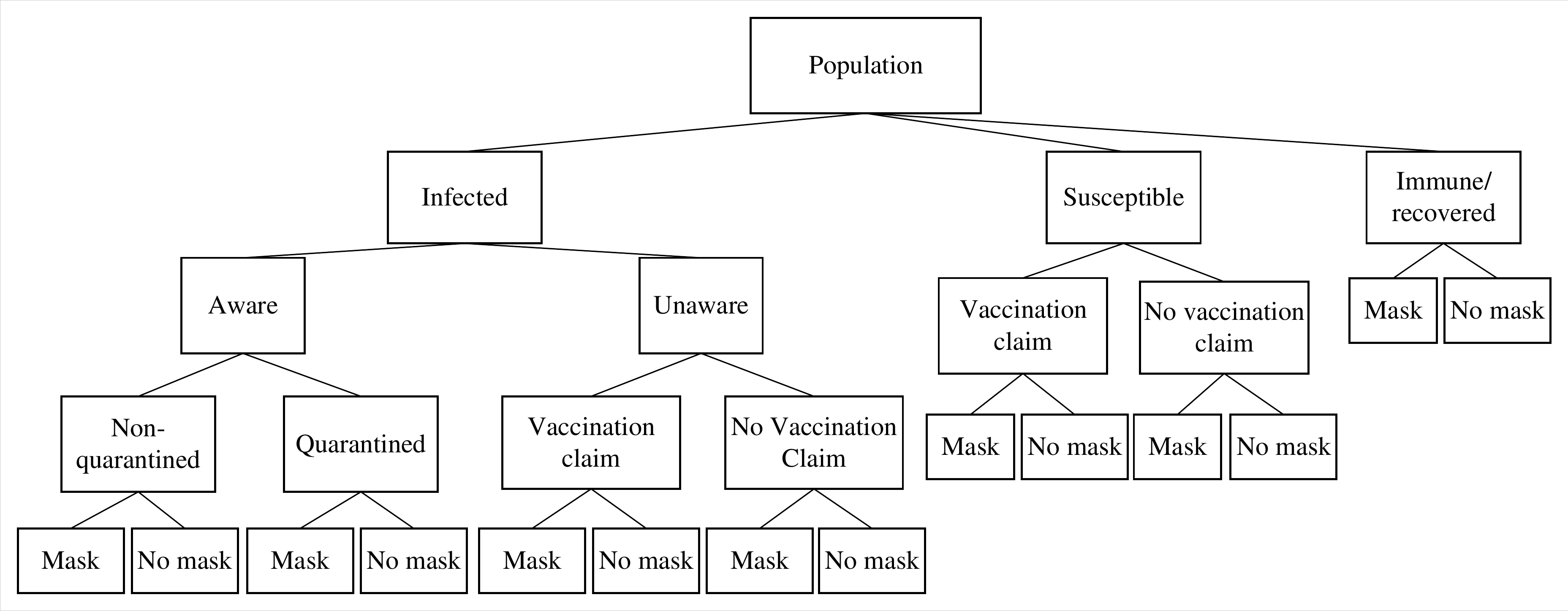}
	\caption{Tree diagram of the factors that determine the highly susceptible, infected, and contagious individuals and the sequence of states of an individual used in the proposed SEIR extended models. The diagram shows the considered states in this paper.}
	\label{fig:treeDiagram}
\end{figure}

The different compartments of the SEIR-V and SEIR-VB models follow the tree diagram shown in Fig.~\ref{fig:treeDiagram}. False vaccination claims might occur with probability $p_{fv}$. Infected individuals are aware of infection with probability $p_a$ and may conceal the infection with probability $p_{ci}$. The percentage of the mask-wearing population and the protection efficacy of mask wearing are denoted by $p_m$ and $m_e$, respectively.

Individuals with an active immunization proof (i.e., a proof showing the period that immunization is active) are allowed to socialize and participate in regular activities. Individuals without the proof may be allowed to socialize with restrictions with probability $p_{sn}$. Table~\ref{tab:notation} lists the notations used in this paper. $S_k$ and $I_k$ for an immunization system without blockchain are denoted as $S_w$ as in (\ref{eq:sw}) and $I_w$ as in (\ref{eq:iw}), and for a system with blockchain, as $S_b$ as in (\ref{eq:sb}) and $I_b$ as in (\ref{eq:ib}), respectively. See the code of these models in \cite{datalink-Medina}.

\begin{table*}[ht]
    \begin{center}
    \small
    \resizebox{1\columnwidth}{!}{
    \begin{tabularx}{0.90\textwidth}{cXc}
    \hline
    \textbf{Variable}&\textbf{Description}&\textbf{Default value}\\
    \hhline{===}
        \multirow {1}*{$\beta$}&Contagious rate per day\\\hline
        $R_{0}$& Basic reproduction number&2.5\\\hline
         $\delta$&Incubation rate&1/5\\\hline
         $\gamma$& Recovery rate&1/9\\\hline
         $\rho$&Fatality rate&1/19 \\\hline
         $\alpha$&Case-fatality rate&2/100\\\hline
         \multirow {1}*{$\lambda$}&Rate of presymptomatic to symptomatic infection& \multirow {1}*{1/14}\\\hline
         \multirow {1}*{$\mu$}&Rate of recovery asymptomatic infection&\multirow {1}*{1/14}\\\hline
         \multirow {1}*{$p_{m}$}& Percentage of the population wearing mask&\\\hline
         $m_{e}$&Mask efficacy&80\\\hline
         $v_{e}$&Vaccine efficacy&95\% \\\hline
         $v_{r}$&Daily vaccination rate&5/100\\\hline
         $v_{i}$&Immunity rate from vaccination&1/14\\\hline
         $l_{v}$&Likelihood of vaccine willingness&8/10 \\\hline
         $p_{is}$&Probability of symptomatic infection&4/10\\\hline
         \multirow {1}*{$p_{ia}$}&Probability of asymptomatic infection&\multirow {1}*{6/10}\\\hline
         \multirow {1}*{$p_{ip}$}&Probability of asymptomatic to symptomatic infection&\multirow {1}*{3/10}\\\hline
         \multirow {1}*{$p_{a}$}&Probability of infected individuals aware of infection&\multirow {1}*{6/10}\\\hline
         \multirow {1}*{$p_{ci}$}&Probability of concealing infection by an infected individual aware of infection&\multirow {1}*{3/10}\\\hline
         \multirow {1}*{$p_{fv}$}&Probability of incomplete or false immunization claim&\multirow {1}*{3/10}\\\hline
         $p_{sn}$&Probability of unsafe socializing&{9/10}\\\hline
         \multirow {1}*{$p_{si}$}&Probability of disease transmission by an infected individual aware of infection&\multirow {1}*{1/10}\\\hhline{===}
    \end{tabularx}
    }
    \end{center}
    \caption{Parameters used in the evaluation of cases and deaths of the SEIR-V and SEIR-VB models.}
    \label{tab:notation}
\end{table*}

\section{RESULTS}
The numerical evaluations of SEIR-V and SEIR-VB show the difference in the number of cases and deaths for a period of 120 days using parameters as reported in the literature (Table~\ref{tab:notation} also shows the default values). Both SEIR-V and SEIR-VB are applied to a population of 330 million individuals (e.g., USA population), as an example. As initial conditions, the population considers 2\% infected cases~\cite{sanchez2021indirect}, which includes symptomatic and asymptomatic cases, and the rest as susceptible. These models were developed during 2020 and 2021.

The evaluation considers the following parameters: A reproduction number of 2.5~\cite{bartsch2020vaccine}; a case fatality rate of 2/100~\cite{he2020estimation}; an average infection, incubation, and critical infection periods of 9, 5, and 18 days, respectively~\cite{cevik2020sars, walsh2020duration,lauer2020incubation,zhou2020clinical}; a probability of infection awareness of 2/100~\cite{hellewell2020feasibility}; a vaccine efficacy of 95\%~\cite{mahase2020covid,polack2020safety}; the start of the vaccination rollout on day 1; a daily vaccination rate of 5/1000~\cite{ourworldindata}; a vaccine willingness of 80\%~\cite{szilagyi2021national,reiter2020acceptability}; and an immunity period of 14 days after vaccination~\cite{baden2021efficacy,euaemergency}.

The models assume that individuals with a vaccination proof socialize without restrictions, while those without a proof may socialize with a 10\% restriction. Immunity may be claimed arbitrarily by any individual. Infected individuals may conceal the infection with 30/100 probability~\cite{levine2010people}, and neglect prevention rules. The probability of asymptomatic infection is set to 60/100, from which 30/100 may become symptomatic after 14 days~\cite{johansson2021sars, white2020asymptomatic, oran2020prevalence}. Both the daily and total reduced number of cases and deaths are analyzed considering different percentages of the population that wear masks. The mask-wearing efficacy is set to 80\%~\cite{howard2021evidence, wang2020reduction}. Table~\ref{tab:notation} shows the default parameters used in the evaluation of the SEIR-V and SEIR-VB models.

\begin{figure}[ht]
	\centering
	\includegraphics[width=\textwidth,trim=5 5 5 5,clip]{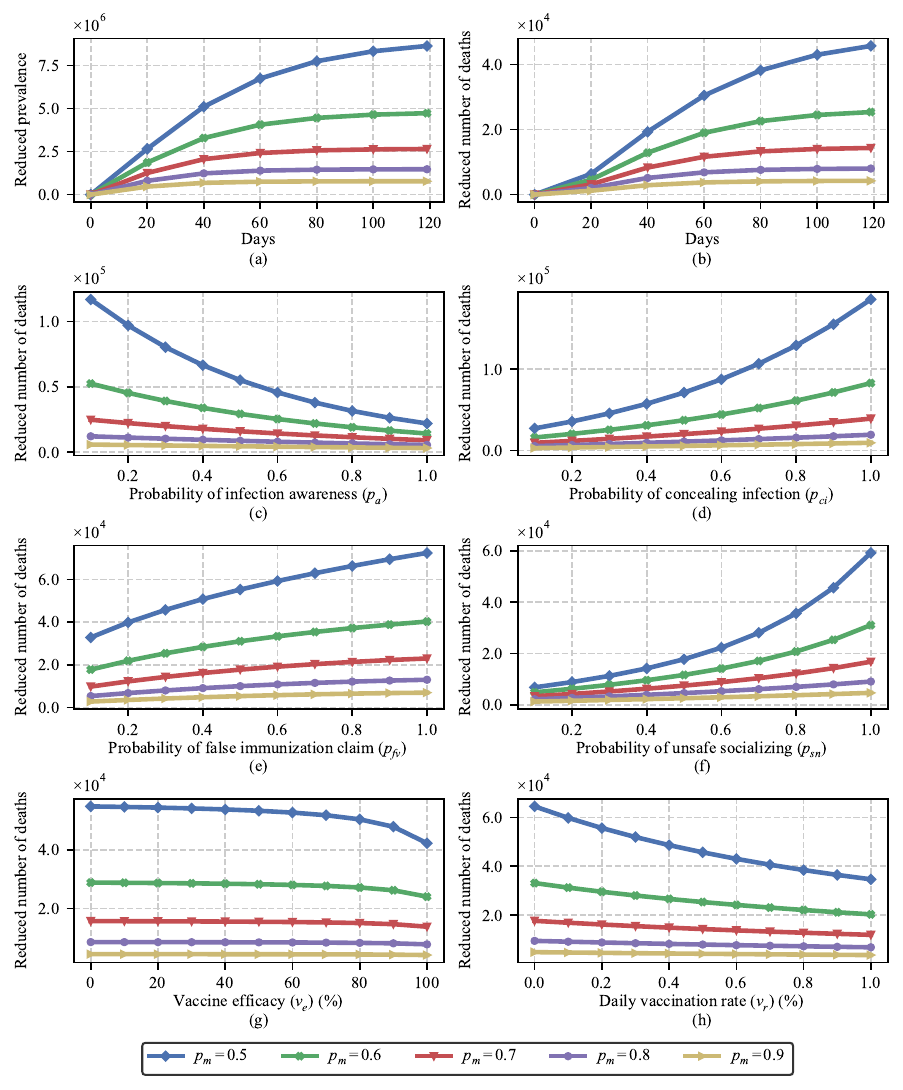}
	\caption{Reduced (a) prevalence; (b) cumulative reduced number of deaths; and total reduced number of deaths using a blockchain-based immunization system for a period of 120 days, as a function of (c) probability of infection awareness; (d) probability of concealing infection; (e) probability of unsafe socializing; (f) vaccine efficacy; and (g) daily vaccination rates. Prevalence and number of deaths are more effectively reduced as conditions worsens. It is in such cases that blockchain is more impactful.}
	\label{fig:group}
\end{figure}


\begin{table*}[ht]
    \begin{center}
    \small
  \renewcommand{\arraystretch}{1.2}
    \begin{tabularx}{0.95\textwidth}{ccr*{5}{Y} }
   \cmidrule{1-8}\morecmidrules\cmidrule{1-8}
    \multicolumn{1}{c}{\multirow{5}{*}{\makecell{\textbf{Basic}\\\textbf{Reproduction}\\\textbf{\makecell{Number\\($R_{0}$)}}}}} & \multicolumn{1}{c}{\multirow{5}{*}{ \makecell{\textbf{Probability}\\\textbf{of unsafe}\\\textbf{ \makecell{Socializing\\\textbf{($p_{sn}$)}}}}}} &
    \multicolumn{6}{c}{\textbf{(Reduced number of cases, Reduced number of deaths) in thousands} } \\
\cline{3-8}
&&\multicolumn{6}{c}{\textbf{Daily Vaccination Rate ($v_r$) (\%)} }\\
\cline{3-8}
&& \multicolumn{3}{c|}{\textbf{Low} (0.10)} & \multicolumn{3}{c}{\textbf{High} (0.50)} \\
\cline{3-8}
&& \multicolumn{6}{c}{\textbf{Percentage of the Population Wearing Masks ($p_m$) (\%)}} \\
\cline{3-8}
& &\textbf{Low} (10)&\textbf{Medium} (50)&\textbf{High} (90)&\textbf{Low} (10)&\textbf{Medium} (50)&\textbf{High} (90)\\
\cline{1-8}
    \multirow{6}{*}{2.5} &\multirow{2}*{0.10}&(4,549.04,&(1,392.31,&(258.99,&(3,904.62,&(1,256.25,&(238.22,\\
        &&24.82)&7.68)&1.43)&21.15)&6.85)&1.30)\\
\cline{2-8}
&\multirow{2}*{0.50}&(28,327.80,&(3,851.26,&(524.70,&(19,614.47,&(3,288.74,&(478.83,\\
&&142.88)&20.99)&2.90)&101.49)&17.81)&2.61)\\
\cline{2-8}
&\multirow{2}*{1.00}&(204,776.90,&(15,566.14,&(963.47,&(149,254.40,&(11,295.44,&(863.80,\\
&&1,028.72)&80.30)&5.31)&742.90)&59.25)&4.71)\\
\cline{1-8}
    \multirow{6}{*}{3.0} &\multirow{2}*{0.10}&(6,245.50,&(1,745.62,&(313.37,
&(5,235.33,&(1565.76,&(287.97,\\
&&33.85)&9.61)&1.73)&28.26)&8.53)&1.57)\\
\cline{2-8}
&\multirow{2}*{0.50}&(52,525.99,&(5,330.80,&(644.18,&(34,066.27,&(4,443.70,&(586.27,\\
&&256.35)&28.87)&3.56)&172.10)&23.97)&3.20)\\
\cline{2-8}
&\multirow{2}*{1.00}&(243,863.00,&(27,648.09,&(1,221.79,&(199,461.10,&(18,734.58,&(1,087.45,\\
&&1,284.53)&138.31)&6.72)&1,029.78)&96.38)&5.92)\\
\cline{1-8}
\multirow{6}{*}{3.5} &\multirow{2}*{0.10}&(8,431.48,&(2,131.15,&(368.64,&(6,877.79,
&(1,899.43,&(338.44,\\
&&45.32)&11.72)&2.04)&36.96)&10.34)&1.85)\\
\cline{2-8}
&\multirow{2}*{0.50}&(89,492.53,&(7,247.59,&(769.17,&(56,559.08,&(5,878.57,&(698.05,\\
&&428.90)&38.92)&4.24)&280.07)&31.57)&3.81))\\
\cline{2-8}
& \multirow{2}*{1.00}&(255,965.10,&(47,324.80,&(1508.76,&(225,759.70,&(30,368.07,&(1,332.48,\\
&&1,389.28)&230.20)&8.29)&1,197.98)&153.03)&7.25)\\
    \cmidrule{1-8}\morecmidrules\cmidrule{1-8}
    \end{tabularx}
\end{center}
\caption{Cases and deaths for different virus reproduction numbers, probability of unsafe socializing, and vaccination rates. Cases and deaths can be decreased by using both blockchain and prevention measures despite large reproduction numbers.}
  \label{tab:results}
\end{table*}

\subsection{Reduced Cases and Deaths by Blockchain}
Fig.~\ref{fig:group} shows the reduced number of cases and deaths by using the blockchain management system. Fig. \ref{fig:group}(a) shows the reduced prevalence of COVID-19 cases through enforcing prevention measures, such as managing access to places that require a reduced social distance by verification of vaccination certificates. Fig. \ref{fig:group}(b) shows the reduced total number of deaths. The figures show that during the first 20 days of the vaccination rollout, the blockchain management system would decrease the number of cases by more than 2.5 million with 50\% of the population wearing masks and about 500,000 cases with 90\% of the population wearing masks. Fig. \ref{fig:group}(b) shows that the blockchain system can reduce the number of deaths by about 46,000 with 50\% of the population wearing masks and about 4,200 deaths with 90\% of the population wearing masks. 

\subsection{Impact by People Behavior and Vaccine Features}
We classify the variables into two groups: prevention measures that decrease the incidence, such as vaccination efficacy, vaccination rate, and social distancing, and negative social behavior that increases the incidence, such as infected individuals who are aware or unaware of their infections and unvaccinated individuals claiming being vaccinated, all ignoring social distancing.

Fig. \ref{fig:group}(c) shows that the number of deaths decreases as individuals become aware of their infection and quarantine themselves. These results show that the adoption of blockchain could reduce more than 30,000 deaths for 70\% of infection awareness with 50\% of the population wearing masks. The difference in the number of deaths increases as the percentage of the wearing-mask population decreases because blockchain enables the detection of high-risk individuals as more individuals are exposed. The blockchain system may reduce about 3,000 deaths with 90\% of the population wearing masks and 100\% of infection awareness.

\subsection{Impact of Individual’s Behavior}
Fig. \ref{fig:group}(d) shows that the blockchain system may reduce about 27,000 deaths with 10\% of the infected individuals concealing their infection and 50\% of the population wearing masks. On the other hand, this system may decrease about 3,000 deaths with 90\% of the population wearing masks. The blockchain system also shows benefits against false vaccination claims.  Fig. \ref{fig:group}(e) shows that the blockchain system may reduce about 70,000 deaths for a 90\% false vaccination claim rate with 50\% of the population wearing masks and about 6,700 deaths with 90\% of the population wearing masks. Its smallest impact is a reduction of about 2,700 deaths for a 10\% rate of false vaccination claims with 90\% of the population wearing masks.

The blockchain management system also reduces the number of cases and deaths when individuals neglect social distancing. It reduces about 60,000 deaths when each individual practices unsafe social distancing with 50\% of the population wearing masks, as shown in Fig. \ref{fig:group}(f). The smallest impact can be seen by a reduction of about 5,000 deaths with 90\% of the population wearing masks. 

\subsection{Impact of Vaccine Features}
Fig. \ref{fig:group}(g) shows that a 100\% effective vaccine alone would be insufficient to avoid as many deaths as the combination of vaccination, mask wearing, and social distancing would. The blockchain system combined with a 100\% effective vaccine would reduce more than 40,000 deaths with 50\% of the population wearing masks and about 4,000 deaths with 90\% of the population wearing masks. A lower efficacy vaccine may increase the number of deaths and that may increase the role importance of the blockchain system.

Fig. \ref{fig:group}(h) shows that at a daily vaccination rate of 1\% of the population, the blockchain system would reduce 3,700 and 35,000 deaths with 90 and 50\% of the population wearing masks, respectively. Table~\ref{tab:results} shows that the blockchain system can reduce the number of cases as the contagion level increases. For example, the proposed system may reduce 15 and 47 million cases under no social distance but with 50\% of the population wearing masks and a low vaccination rate of 0.1\%, for virus reproduction numbers of 2.5 and 3.5, respectively. Also, the blockchain system could reduce 1.3 and 1.9 million cases for a high vaccination rate of 0.5\%, with 50\% of the population wearing masks, and individuals socializing with 90\% restrictions with reproduction numbers of 2.5 and 3.5, respectively.

With the vaccination rollout, the number of individuals wearing masks may decrease but that could increase the number of cases and deaths. Table~\ref{tab:results} shows that with a low daily vaccination rate of 0.1\%, with 50\% of the population wearing masks and no individuals keeping social distance, the system could reduce more than 80,000 deaths for a (virus) reproduction number of 2.5 and more than 230,000 deaths for a reproduction number of 3.5. On the other hand, at a daily vaccination rate of 0.5\%, 90\% of the population wearing masks, and individuals socializing with 90\% restrictions, the system could still reduce 1,300 and 1,850 deaths for reproduction numbers of 2.5 and 3.5, respectively. Even though these numbers of deaths are small in comparison with the initial population, decreasing incidence and mortality is deemed significant.

\section{DISCUSSION}
Modeling and forecasting the incidence of infectious diseases such a COVID-19 is challenging because of the dynamics of the many involved parameters~\cite{bertozzi2020challenges}. The parameters used in our evaluations are time invariant. However, as more people become vaccinated, mask wearing rates may decrease and evaluations of such scenario would be needed. 

Using blockchain to track vaccinated individuals would support the enforcing of prevention measures. These evaluations consider a uniform enforcement of preventive rules across the whole population in the SEIR-VB model, but non-uniformity may be more realistic in a country and that may need to be considered. Enforcing the prevention measures in the world has been challenging. Therefore, the use of blockchain may face some resistance and a flexible approach may need to be developed. Also, the current rollout may target fast vaccination of the population to provide a fast health and economic relief. Such a case may include variable vaccination rates that may need be analyzed in similar performance terms as done in this paper.

\section{CONCLUSION}

Blockchain provides a means to reliably record vaccinations and issue time-dependent certificates that can be used to enforce vaccination verification in a pandemic. We modeled and evaluated a vaccine distribution management system that reliably verifies whether an individual is vaccinated so that such individual can be allowed to participate in social activities of normal daily life, including reduced social distancing, traveling, or group gatherings.

The performance of the proposed blockchain system was evaluated by the reduced number of cases and deaths between the blockchain management system and a conventional system that uses no blockchain. The results indicate that the adoption of blockchain as an immunization control system could reduce the number of cases and deaths, as an important effect of performing preventive measures.


\clearpage
\bibliographystyle{unsrt}  
\bibliography{references}

\begin{thebibliography}{10}

\bibitem{velavan2020covid}
Thirumalaisamy~P Velavan and Christian~G Meyer.
\newblock {The COVID-19 epidemic}.
\newblock {\em Tropical medicine \& international health}, 25(3):278, 2020.

\bibitem{wu2020sars}
Di~Wu, Tiantian Wu, Qun Liu, and Zhicong Yang.
\newblock {The SARS-CoV-2 outbreak: what we know}.
\newblock {\em International Journal of Infectious Diseases}, 94:44--48, 2020.

\bibitem{pedersen2020sars}
Savannah~F Pedersen, Ya-Chi Ho, et~al.
\newblock {SARS-CoV-2: a storm is raging}.
\newblock {\em The Journal of clinical investigation}, 130(5), 2020.

\bibitem{fathi2020time}
Sina Fathi-Kazerooni, Roberto Rojas-Cessa, Ziqian Dong, and Vatcharapan
  Umpaichitra.
\newblock Time series analysis and correlation of subway turnstile usage and
  covid-19 prevalence in new york city.
\newblock {\em arXiv preprint arXiv:2008.08156}, 2020.

\bibitem{fathi2021correlation}
Sina Fathi-Kazerooni, Roberto Rojas-Cessa, Ziqian Dong, and Vatcharapan
  Umpaichitra.
\newblock Correlation of subway turnstile entries and covid-19 incidence and
  deaths in new york city.
\newblock {\em Infectious Disease Modelling}, 6:183--194, 2021.

\bibitem{piraveenan2020optimal}
Mahendra Piraveenan, Shailendra Sawleshwarkar, Michael Walsh, Iryna Zablotska,
  Samit Bhattacharyya, Habib~Hassan Farooqui, Tarun Bhatnagar, Anup Karan,
  Manoj Murhekar, Sanjay Zodpey, et~al.
\newblock Optimal governance and implementation of vaccination programs to
  contain the covid-19 pandemic.
\newblock {\em arXiv preprint arXiv:2011.06455}, 2020.

\bibitem{krueger2021risk}
Tyll Krueger, Krzysztof Gogolewski, Marcin Bodych, Anna Gambin, Giulia
  Giordano, Sarah Cuschieri, Thomas Czypionka, Matjaz Perc, Elena Petelos,
  Magdalena Rosi{\'n}ska, et~al.
\newblock Risk of covid-19 epidemic resurgence with the introduction of
  vaccination passes.
\newblock {\em medRxiv. The Preprint Server for Health Sciences}, 2021.

\bibitem{vaccines2020randomized}
Janssen Vaccines.
\newblock {A Randomized, Double-blind, Placebo-controlled Phase 3 Study to
  Assess the Efficacy and Safety of Ad26.COV2.S for the Prevention of
  SARS-CoV-2-mediated COVID-19 in Adults Aged 18 Years and Older}, 2020.

\bibitem{burki2020russian}
Talha~Khan Burki.
\newblock {The Russian vaccine for COVID-19}.
\newblock {\em The Lancet Respiratory Medicine}, 8(11):e85--e86, 2020.

\bibitem{knoll2021oxford}
Maria~Deloria Knoll and Chizoba Wonodi.
\newblock {Oxford--AstraZeneca COVID-19 vaccine efficacy}.
\newblock {\em The Lancet}, 397(10269):72--74, 2021.

\bibitem{priesemann2021action}
Viola Priesemann, Rudi Balling, Melanie~M Brinkmann, Sandra Ciesek, Thomas
  Czypionka, Isabella Eckerle, Giulia Giordano, Claudia Hanson, Zdenek Hel,
  Pirta Hotulainen, et~al.
\newblock An action plan for pan-european defence against new sars-cov-2
  variants.
\newblock {\em The Lancet}, 397(10273):469--470, 2021.

\bibitem{mcghin2019blockchain}
Thomas McGhin, Kim-Kwang~Raymond Choo, Charles~Zhechao Liu, and Debiao He.
\newblock {Blockchain in healthcare applications: Research challenges and
  opportunities}.
\newblock {\em Journal of Network and Computer Applications}, 135:62--75, 2019.

\bibitem{ahmad2020blockchain}
Raja~Wasim Ahmad, Khaled Salah, Raja Jayaraman, Ibrar Yaqoob, Samer Ellahham,
  and Mohammed Omar.
\newblock {Blockchain and COVID-19 Pandemic: Applications and Challenges}.

\bibitem{bansal2020optimizing}
Agam Bansal, Chandan Garg, and Rana~P Padappayil.
\newblock {Optimizing the Implementation of COVID-19 “Immunity
  Certificates” Using Blockchain}.
\newblock {\em Journal of Medical Systems}, 44(9):1--2, 2020.

\bibitem{eisenstadt2020covid}
Marc Eisenstadt, Manoharan Ramachandran, Niaz Chowdhury, Allan Third, and John
  Domingue.
\newblock {COVID-19 Antibody Test/Vaccination Certification: There’s an App
  for That}.
\newblock {\em IEEE Open Journal of Engineering in Medicine and Biology},
  1:148--155, 2020.

\bibitem{hasan2020blockchain}
Haya~R Hasan, Khaled Salah, Raja Jayaraman, Junaid Arshad, Ibrar Yaqoob,
  Mohammed Omar, and Samer Ellahham.
\newblock {Blockchain-Based Solution for COVID-19 Digital Medical Passports and
  Immunity Certificates }.
\newblock {\em IEEE Access}, 2020.

\bibitem{nakamoto2019bitcoin}
Satoshi Nakamoto.
\newblock {Bitcoin: A Peer-to-Peer Electronic Cash System}.
\newblock Technical report, Manubot, 2019.

\bibitem{crosby2016blockchain}
Michael Crosby, Pradan Pattanayak, Sanjeev Verma, Vignesh Kalyanaraman, et~al.
\newblock {Blockchain technology: Beyond Bitcoin}.
\newblock {\em Applied Innovation}, 2(6-10):71, 2016.

\bibitem{angraal2017blockchain}
Suveen Angraal, Harlan~M Krumholz, and Wade~L Schulz.
\newblock {Blockchain Technology: Applications in Health Care}.
\newblock {\em Circulation: Cardiovascular quality and outcomes},
  10(9):e003800, 2017.

\bibitem{engelhardt2017hitching}
Mark~A Engelhardt.
\newblock {Hitching Healthcare to the Chain: An Introduction to Blockchain
  Technology in the Healthcare Sector}.
\newblock {\em Technology Innovation Management Review}, 7(10), 2017.

\bibitem{linn2016blockchain}
Laure~A Linn and Martha~B Koo.
\newblock {Blockchain For Health Data and Its Potential Use in Health IT and
  Health Care Related Research }.
\newblock In {\em ONC/NIST Use of Blockchain for Healthcare and Research
  Workshop. Gaithersburg, Maryland, United States: ONC/NIST}, pages 1--10,
  2016.

\bibitem{8643084}
S.~{Wang}, L.~{Ouyang}, Y.~{Yuan}, X.~{Ni}, X.~{Han}, and F.~{Wang}.
\newblock {Blockchain-Enabled Smart Contracts: Architecture, Applications, and
  Future Trends}.
\newblock {\em IEEE Transactions on Systems, Man, and Cybernetics: Systems},
  49(11):2266--2277, 2019.

\bibitem{marbouh2020blockchain}
Dounia Marbouh, Tayaba Abbasi, Fatema Maasmi, Ilhaam~A Omar, Mazin~S Debe,
  Khaled Salah, Raja Jayaraman, and Samer Ellahham.
\newblock {Blockchain for COVID-19: Review, Opportunities, and a Trusted
  Tracking System}.
\newblock {\em Arabian Journal for Science and Engineering}, pages 1--17, 2020.

\bibitem{nguyen2020blockchain}
Dinh Nguyen, Ming Ding, Pubudu~N Pathirana, and Aruna Seneviratne.
\newblock {Blockchain and AI-based Solutions to Combat Coronavirus
  (COVID-19)-like Epidemics: A Survey}.
\newblock 2020.

\bibitem{tseng2018governance}
Jen-Hung Tseng, Yen-Chih Liao, Bin Chong, and Shih-wei Liao.
\newblock {Governance on the Drug Supply Chain via Gcoin Blockchain}.
\newblock {\em International journal of environmental research and public
  health}, 15(6):1055, 2018.

\bibitem{yong2020intelligent}
Binbin Yong, Jun Shen, Xin Liu, Fucun Li, Huaming Chen, and Qingguo Zhou.
\newblock {An intelligent blockchain-based system for safe vaccine supply and
  supervision}.
\newblock {\em International Journal of Information Management}, 52:102024,
  2020.

\bibitem{li1995global}
Michael~Y Li and James~S Muldowney.
\newblock {GLOBAL STABILITY FOR THE SEIR MODEL IN EPIDEMIOLOGY}.
\newblock {\em Mathematical biosciences}, 125(2):155--164, 1995.

\bibitem{delamater2019complexity}
Paul~L Delamater, Erica~J Street, Timothy~F Leslie, Y~Tony Yang, and Kathryn~H
  Jacobsen.
\newblock {Complexity of the Basic Reproduction Number (R0)}.
\newblock {\em Emerging infectious diseases}, 25(1):1, 2019.

\bibitem{lee2020clinical}
Seungjae Lee, Tark Kim, Eunjung Lee, Cheolgu Lee, Hojung Kim, Heejeong Rhee,
  Se~Yoon Park, Hyo-Ju Son, Shinae Yu, Jung~Wan Park, et~al.
\newblock {Clinical Course and Molecular Viral Shedding Among Asymptomatic and
  Symptomatic Patients With SARS-CoV-2 Infection in a Community Treatment
  Center in the Republic of Korea}.
\newblock {\em JAMA internal medicine}, 180(11):1447--1452, 2020.

\bibitem{datalink-Medina}
Jorge Medina and Roberto Rojas-Cessa.
\newblock {Model Code of Extended SEIR models with Vaccination Rollout and
  Blockchain}, 2021.

\bibitem{sanchez2021indirect}
Miguel S{\'a}nchez-Romero, Vanessa di~Lego, Alexia Prskawetz, and Bernardo
  L.~Queiroz.
\newblock An indirect method to monitor the fraction of people ever infected
  with covid-19: An application to the united states.
\newblock {\em PloS one}, 16(1):e0245845, 2021.

\bibitem{bartsch2020vaccine}
Sarah~M Bartsch, Kelly~J O'Shea, Marie~C Ferguson, Maria~Elena Bottazzi,
  Patrick~T Wedlock, Ulrich Strych, James~A McKinnell, Sheryl~S Siegmund,
  Sarah~N Cox, Peter~J Hotez, et~al.
\newblock {Vaccine Efficacy Needed for a COVID-19 Coronavirus Vaccine to
  Prevent or Stop an Epidemic as the Sole Intervention}.
\newblock {\em American journal of preventive medicine}, 59(4):493--503, 2020.

\bibitem{he2020estimation}
Wenqing He, Grace~Y Yi, and Yayuan Zhu.
\newblock {Estimation of the basic reproduction number, average incubation
  time, asymptomatic infection rate, and case fatality rate for COVID-19:
  Meta-analysis and sensitivity analysis}.
\newblock {\em Journal of medical virology}, 92(11):2543--2550, 2020.

\bibitem{cevik2020sars}
Muge Cevik, Matthew Tate, Ollie Lloyd, Alberto~Enrico Maraolo, Jenna Schafers,
  and Antonia Ho.
\newblock {SARS-CoV-2, SARS-CoV, and MERS-CoV viral load dynamics, duration of
  viral shedding, and infectiousness: a systematic review and meta-analysis}.
\newblock {\em The Lancet Microbe}, 2020.

\bibitem{walsh2020duration}
Kieran~A Walsh, Susan Spillane, Laura Comber, Karen Cardwell, Patricia
  Harrington, Jeff Connell, Conor Teljeur, Natasha Broderick, Cillian~F
  de~Gascun, Susan~M Smith, et~al.
\newblock {The duration of infectiousness of individuals infected with
  SARS-CoV-2}.
\newblock {\em Journal of Infection}, 2020.

\bibitem{lauer2020incubation}
Stephen~A Lauer, Kyra~H Grantz, Qifang Bi, Forrest~K Jones, Qulu Zheng,
  Hannah~R Meredith, Andrew~S Azman, Nicholas~G Reich, and Justin Lessler.
\newblock {The Incubation Period of Coronavirus Disease 2019 (COVID-19) From
  Publicly Reported Confirmed Cases: Estimation and Application}.
\newblock {\em Annals of internal medicine}, 172(9):577--582, 2020.

\bibitem{zhou2020clinical}
Fei Zhou, Ting Yu, Ronghui Du, Guohui Fan, Ying Liu, Zhibo Liu, Jie Xiang,
  Yeming Wang, Bin Song, Xiaoying Gu, et~al.
\newblock {Clinical course and risk factors for mortality of adult inpatients
  with COVID-19 in Wuhan, China: a retrospective cohort study}.
\newblock {\em The lancet}, 395(10229):1054--1062, 2020.

\bibitem{hellewell2020feasibility}
Joel Hellewell, Sam Abbott, Amy Gimma, Nikos~I Bosse, Christopher~I Jarvis,
  Timothy~W Russell, James~D Munday, Adam~J Kucharski, W~John Edmunds, Fiona
  Sun, et~al.
\newblock {Feasibility of controlling COVID-19 outbreaks by isolation of cases
  and contacts}.
\newblock {\em The Lancet Global Health}, 8(4):e488--e496, 2020.

\bibitem{mahase2020covid}
Elisabeth Mahase.
\newblock {Covid-19: Moderna vaccine is nearly 95\% effective, trial involving
  high risk and elderly people shows}.
\newblock {\em BMJ: British Medical Journal (Online)}, 371, 2020.

\bibitem{polack2020safety}
Fernando~P Polack, Stephen~J Thomas, Nicholas Kitchin, Judith Absalon,
  Alejandra Gurtman, Stephen Lockhart, John~L Perez, Gonzalo P{\'e}rez~Marc,
  Edson~D Moreira, Cristiano Zerbini, et~al.
\newblock {Safety and Efficacy of the BNT162b2 mRNA Covid-19 Vaccine}.
\newblock {\em New England Journal of Medicine}, 383(27):2603--2615, 2020.

\bibitem{ourworldindata}
Coronavirus (covid-19) vaccinations.
\newblock Our World in Data, April 6, 2021.

\bibitem{szilagyi2021national}
Peter~G Szilagyi, Kyla Thomas, Megha~D Shah, Nathalie Vizueta, Yan Cui, Sitaram
  Vangala, and Arie Kapteyn.
\newblock {National Trends in the US Public’s Likelihood of Getting a
  COVID-19 Vaccine—April 1 to December 8, 2020}.
\newblock {\em JAMA}, 325(4):396--398, 2021.

\bibitem{reiter2020acceptability}
Paul~L Reiter, Michael~L Pennell, and Mira~L Katz.
\newblock {Acceptability of a COVID-19 vaccine among adults in the United
  States: How many people would get vaccinated?}
\newblock {\em Vaccine}, 38(42):6500--6507, 2020.

\bibitem{baden2021efficacy}
Lindsey~R Baden, Hana~M El~Sahly, Brandon Essink, Karen Kotloff, Sharon Frey,
  Rick Novak, David Diemert, Stephen~A Spector, Nadine Rouphael, C~Buddy
  Creech, et~al.
\newblock Efficacy and safety of the mrna-1273 sars-cov-2 vaccine.
\newblock {\em New England Journal of Medicine}, 384(5):403--416, 2021.

\bibitem{euaemergency}
Application~Type EUA and Sudhakar Agnihothram.
\newblock Emergency use authorization (eua) for an unapproved product review
  memorandum identifying information.

\bibitem{levine2010people}
Timothy~R Levine, Rachel~K Kim, and Lauren~M Hamel.
\newblock {People Lie for a Reason: Three Experiments Documenting the Principle
  of Veracity}.
\newblock {\em Communication Research Reports}, 27(4):271--285, 2010.

\bibitem{johansson2021sars}
Michael~A Johansson, Talia~M Quandelacy, Sarah Kada, Pragati~Venkata Prasad,
  Molly Steele, John~T Brooks, Rachel~B Slayton, Matthew Biggerstaff, and Jay~C
  Butler.
\newblock {SARS-CoV-2 Transmission From People Without COVID-19 Symptoms}.
\newblock {\em JAMA network open}, 4(1):e2035057--e2035057, 2021.

\bibitem{white2020asymptomatic}
Elizabeth~M White, Christopher~M Santostefano, Richard~A Feifer, Cyrus~M Kosar,
  Carolyn Blackman, Stefan Gravenstein, and Vincent Mor.
\newblock {Asymptomatic and Presymptomatic Severe Acute Respiratory Syndrome
  Coronavirus 2 Infection Rates in a Multistate Sample of Skilled Nursing
  Facilities}.
\newblock {\em JAMA Internal Medicine}, 180(12):1709--1711, 2020.

\bibitem{oran2020prevalence}
Daniel~P Oran and Eric~J Topol.
\newblock {Prevalence of Asymptomatic SARS-CoV-2 Infection: A Narrative
  Review}.
\newblock {\em Annals of internal medicine}, 173(5):362--367, 2020.

\bibitem{howard2021evidence}
Jeremy Howard, Austin Huang, Zhiyuan Li, Zeynep Tufekci, Vladimir Zdimal,
  Helene-Mari van~der Westhuizen, Arne von Delft, Amy Price, Lex Fridman,
  Lei-Han Tang, et~al.
\newblock {An evidence review of face masks against COVID-19}.
\newblock {\em Proceedings of the National Academy of Sciences}, 118(4), 2021.

\bibitem{wang2020reduction}
Yu~Wang, Huaiyu Tian, Li~Zhang, Man Zhang, Dandan Guo, Wenting Wu, Xingxing
  Zhang, Ge~Lin Kan, Lei Jia, Da~Huo, et~al.
\newblock {Reduction of secondary transmission of SARS-CoV-2 in households by
  face mask use, disinfection and social distancing: a cohort study in Beijing,
  China}.
\newblock {\em BMJ global health}, 5(5):e002794, 2020.

\bibitem{bertozzi2020challenges}
Andrea~L Bertozzi, Elisa Franco, George Mohler, Martin~B Short, and Daniel
  Sledge.
\newblock {The challenges of modeling and forecasting the spread of COVID-19}.
\newblock {\em Proceedings of the National Academy of Sciences},
  117(29):16732--16738, 2020.

\end{thebibliography}

\end{document}